\documentclass[prints]{aa}
\usepackage{psfig}
\usepackage{graphicx}
\usepackage{supertabular}
\begin{document}


\title{The 2MASS Tully-Fisher relation for flat edge-on galaxies}
\titlerunning{The 2MASS TF relation for edge-on galaxies}

\author{I.D.Karachentsev\inst{1}\and S.N.Mitronova \inst{1}
\and V.E.Karachentseva \inst{2}
\and Yu.N.Kudrya \inst{2}
\and T.H.Jarrett \inst{3}}
\institute{Special Astrophysical Observatory, Russian Academy
   of Sciences, N.Arkhyz, KChR, 369167, Russia
\and Astronomical Observatory of the Kiev National University, 04053,
  Observatorna 3, Kiev, Ukraine
\and Infrared Processing and Analysis Center, Mail Stop 100-22, California
Institute of Technology, Jet Propulsion Laboratory, Pasadena, CA 91125 }
\date{submitted: 10.09.2002; accepted}

\abstract{ The spiral edge-on galaxies from the `Revised Flat Galaxy
Catalog' (RFGC) are identified with the Extended Source Catalog of the
Two Micron All Sky Survey (2MASS). A relative number of the 2MASS
detected galaxies is 2996/4236 = 0.71. We consider the statistical
properties of the Tully-Fisher relations for the edge-on galaxies in
the $B, I, J, H,$ and $K_s$ bands. The slope of derived TF relations
increases steadily from 4.9 in the $B$ band to 9.3 in the $K$ one. The effect
is mainly due to the internal extinction which is different in dwarf
and giant spiral galaxies seen edge-on that leads to the tight correlation
between galaxy color and luminosity. The moderate scatter of the RFGC
galaxies in the ``color-luminosity'' diagram, 0$\fm$86, provides us with
a ``cheap'' method of mass measurements of distances to galaxies on
the basis of modern photometric sky surveys.
}
\maketitle
\section{Introduction}

The Tully-Fisher (TF) relation (Tully \& Fisher,1977) between luminosity
and rotation velocity of spiral galaxies is a basic tool for studying
large-scale motions of galaxies because it provides us with distances
independent of galaxy redshifts. Using this relation, Tully \& Fisher
and their numerous followers excluded usually very tilted spiral
galaxies burdened with strong internal extinction. However, as it was
shown by Karachentsev (1989), internal extinction is not a dominant
reason for scatter of spiral galaxies on the TF diagram, and edge-on
galaxies may be successfully used to map cosmic streamings. For such a
purpose the `Flat Galaxy Catalog', FGC, (Karachentsev et al, 1993)
and its renewed version, RFGC (Karachentsev et al, 1999) were prepeared.
Selection of galaxies into the RFGC was carried out based on two simple
geometric criteria, when the major angular diameter of galaxies is greater
than 0.6 arcmin and the apparent axial ratio, $a/b$, is greater than 7.
Here the major and minor diameters correspond to the standard isophote
of $25^m/\sq\arcsec$ in the $B$ band. The RFGC catalog covers the
entire northern and southern sky, and contains 4236 galaxies mostly
of Sc--Sd morphological types. The all-sky distribution of RFGC galaxies
looks quite smooth because late-type spiral galaxies are mostly found
in the general field, avoiding rich cluster cores with their
large virial motions. Only a small number of the RFGC galaxies have
apparent integral magnitudes, $B_t$, from RC3 (de Vaucouleurs et al. 1991).
But for all its galaxies the RFGC presents marginal B- magnitudes
derived from angular diameters of the galaxies as well as from their
surface brightness and other parameters (Kudrya et al. 1997). Being
reduced to the RC3 photometric system, these magnitudes are characterized
with an error of $\sigma(B) = 0\fm3$. The TF relation `absolute magnitude
vs. HI line width', plotted for $\sim$800 RFGC galaxies, has a slope of $-$5.3
and a scatter of $\sigma(M_B) = 0\fm56$.

\section{ 2MASS photometry of the RFGC galaxies}

  The Two Micron All-Sky Survey (2MASS) was carried out in three near-
infrared bands: $J$ (1.11 - 1.36 $\mu$), $H$ (1.50 - 1.80 $\mu$), and $K_s$
(2.00 - 2.32 $\mu$) using two 1.3-m telescopes in Arizona and Chile, each
designed with three 256$\times$256 pixel NICMOS arrays giving a resolution of
2$\arcsec$/pixel. Over the whole sky the 2MASS led to detection of $\sim$3~000~000
galaxies brighter than $K_s = 14\fm5$ (Jarrett, 2000). About 1.65
million galaxies with $K_s < 14^m$ and an angular diameters greater than
10$\arcsec$ are included in the 2MASS Extended Sources Catalog (XSC). Selection
of extended sources and their photometry was made with the standard set
of algorithms (Jarrett et al. 2000). Photometric calibration of the data was
described by Nikolaev et al. (2000). As a result, the 2MASS XSC contains
a lot of geometric and photometric characteristics of galaxies. Among them
we use the following key parameters in the present paper:

$r_{20}$  ---   major isophotal radius in arcsec, measured at the $20^m/\sq\arcsec$
	  level in the $K_s$ band via photometry in elliptical isophotes;

$<b/a>_s$ --- minor-to-major axial ratio fit to 3$\sigma$ isophote co-added in
	  three $J,H,K$ bands;

$J_{20}, H_{20}, K_{20}$  --- isophotal fiducial elliptical-aperture magnitudes in
	    corresponding bands, measured at the $K=20^m/\sq\arcsec$ level;

$J_{ext}, H_{ext}, K_{ext}$  --- integral "total" magnitudes as derived
	    from the isophotal magnitudes ($J_{20}, H_{20}, K_{20}$) and
	    the extrapolation of the fit to the radial surface brightness
	    distribution. The extrapolation ($r_{ext}$) is carried out to
	    roughly four times the disk scale length. Details are given in
	    Jarrett et al. 2000;

  We carried out mutual identification of objects from RFGC and 2MASS XSC,
taking into account coordinates, as well as dimensions and orientation of
the galaxies. As it has been noted by Jarrett (2000), 2MASS is not sensitive
to late-type blue galaxies, especially of low surface brightness due to the
high background of the NIR sky and the relatively short exposure times
($\sim8~$sec/object). When questionable cases were omitted,
we identified in 2MASS XSC 2996 RFGC galaxies from their
total number of 4236. Therefore, a relative number of the 2MASS
detected galaxies is 71\%.

\begin{figure}
\centering
\includegraphics[width=6.2cm,angle=-90]{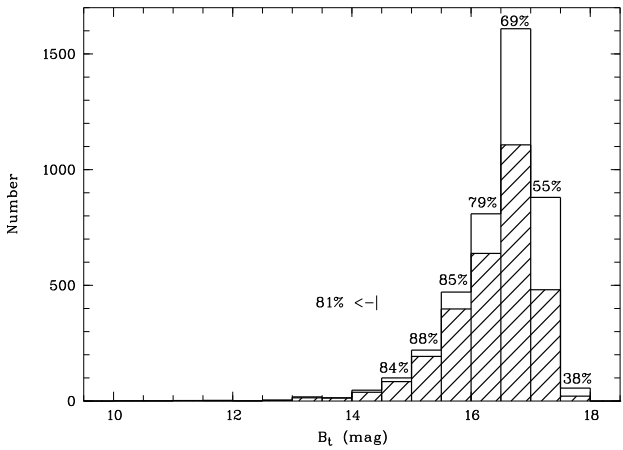}
\vspace{5mm}
\caption{ Apparent blue magnitude distribution for the RFGC galaxies. The
galaxies detected in the 2MASS are shaded.}
\end{figure}

\begin{figure}
\centering
\includegraphics[width=6.2cm,angle=-90]{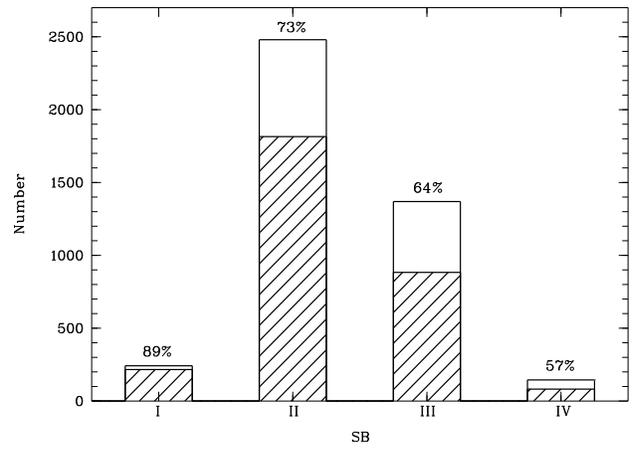}
\vspace{5mm}
\caption{ Distribution of the RFGC galaxies by optical surface brightness
classes from high (I) to very low (IV). The galaxies seen in the 2MASS are
shaded.}
\end{figure}

  In Fig.1 we show the histogram of the distribution of RFGC galaxies according
to their integral blue apparent magnitudes. The galaxies detected in 2MASS
are shaded. As it was to be expected, the 2MASS detection rate decreases towards
faint objects. In RFGC the galaxies were divided into four classes of
their blue surface brightness: I --- high, II --- normal, III --- low, and IV ---
very low. Fig.2 displays the distribution by these classes. The relative
number of galaxies detected in 2MASS (shaded) decreases steadily with
decreasing average optical surface brightness, which obviously comes up
to expectations. A morphological classification of spiral galaxies
seen edge-on can be done reliably based on their bulge-to-disk dimension
ratio. The distribution of RFGC galaxies by morphological types in the
de Vaucouleurs scheme ( 2=Sab, 3=Sb,..., 9=Sm) is presented in Fig.3,

\begin{figure}
\centering
\includegraphics[width=6.2cm,angle=-90]{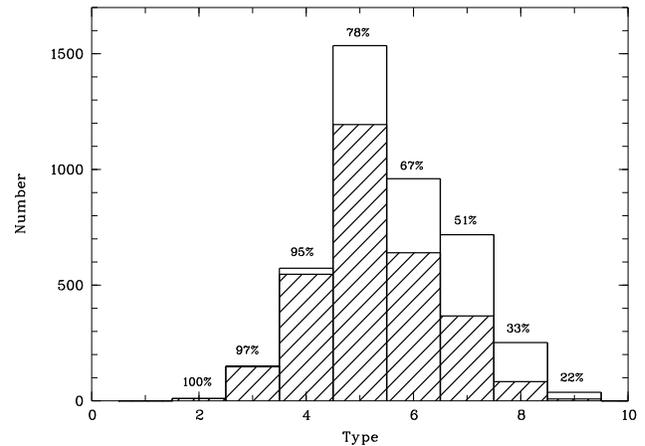}
\vspace{5mm}
\caption{ Morphological type distribution for the RFGC galaxies from Sab=2
to Sm=9. The galaxies detected in the 2MASS are shaded.}
\end{figure}

\begin{figure}
\centering
\includegraphics[width=6.2cm,angle=-90]{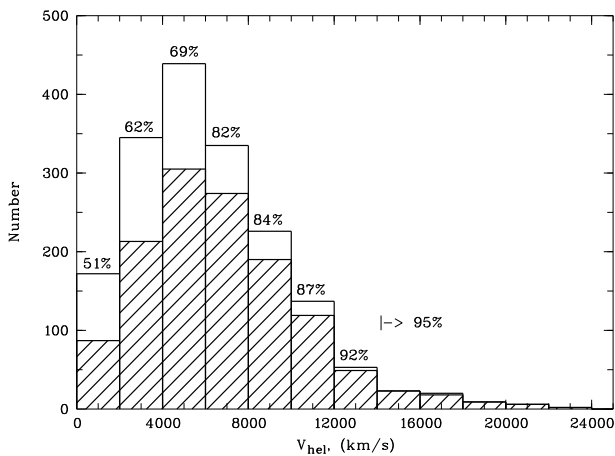}
\vspace{5mm}
\caption{ Radial velocity distribution for the RFGC galaxies. Percentage
of the 2MASS detected objects (shaded) is shown in each velocity bin.}
\end{figure}

where the 2MASS detected galaxies are shaded. As clearly shown, the
fraction of 2MASS detected galaxies decreases along the Hubble sequence
from 100\% for Sab type till 22\% for the latest types. The presence of
such morphological selectivity leads to a paradoxical effect displayed
in Fig.4, where the distribution of 1772 RFGC galaxies by their radial
velocities is shown. Unexpectedly, the relative number of 2MASS detected
sources (shaded) increases steadily from nearby objects towards distant
ones, reaching $\sim$95\%, when $V_h >$ 14000 km/s. These data give us a hint,
that the color of edge-on galaxies, say B--K, may be a sensitive indicator
of their luminosity (see below).

 Due to the high brightness of the NIR sky and short 2MASS exposures,
a bluish periphery of the disks of spiral galaxies is usually unseen above
the $K = 20^m/\sq\arcsec$ isophote. On the average, the infra-red
diameters turn out to be twice as small as the optical ones
Jarrett et al. (2002). Because of bulges of spiral galaxies are  redder than
disks, their infra-red axial ratios, $a/b$, are systematically lower
in 2MASS with respect to the $B$ band. The blue axial ratios for RFGC
galaxies occupy a range of [7 -- 21] with a median of 8.6, while the
infra-red ones cover an interval of [1 -- 10] with a median of 4.1.

 \section{Comparison of 2MASS photometry with the deep I photometry}.

 Plotting the TF diagram, one usually uses the total, but not isophotal
magnitude of galaxies. In the 2MASS XSC the total magnitudes were computed
from the isophotal magnitudes and the radial fit to the surface brightness
profile. The aperture correction, $J_{20} - J_{ext}$, tends to increase
from bright galaxies towards faint ones. The mean aperture correction is
$0\fm24$, i.e. on the average the isophote magnitude underestimates $\sim20$\%
of the total flux. About the same correction is the case for $H$ and $K$
magnitudes in XSC. But the mean value and the dispersion of the aperture
correction depend both on the morphological type and surface brightness
of galaxies. Although the employment of extrapolated apparent magnitudes
seems to be more motivated for plotting the TF relation, we prefer to
deal with the isophotal magnitudes. The reason for the preference
is the uncertainty in the aperture correction for galaxies whose brightness
profile is determined unreliably. In the upper panel of Fig.5 we show the
distribution of RFGC galaxies according to their color $(J_{20} - K_{20})$
and apparent magnitude $K_{20}$. The mean color, $<J_{20} - K_{20}> = 1.21$,
is almost the same for both bright and faint galaxies. The lower panel
presents a similar diagram for the extrapolated (total) magnitudes. The mean
corrected color of galaxies becomes a little bluer, $<J_{ext} - K_{ext}>$
= 1.16, the RMS scatter increases from 0$\fm$16 to 0$\fm$23, and in the diagram there
appear galaxies with colors less than 0.5 and more than 1.8, not typical
of spiral galaxies (Jarrett, 2000).

\begin{figure*}
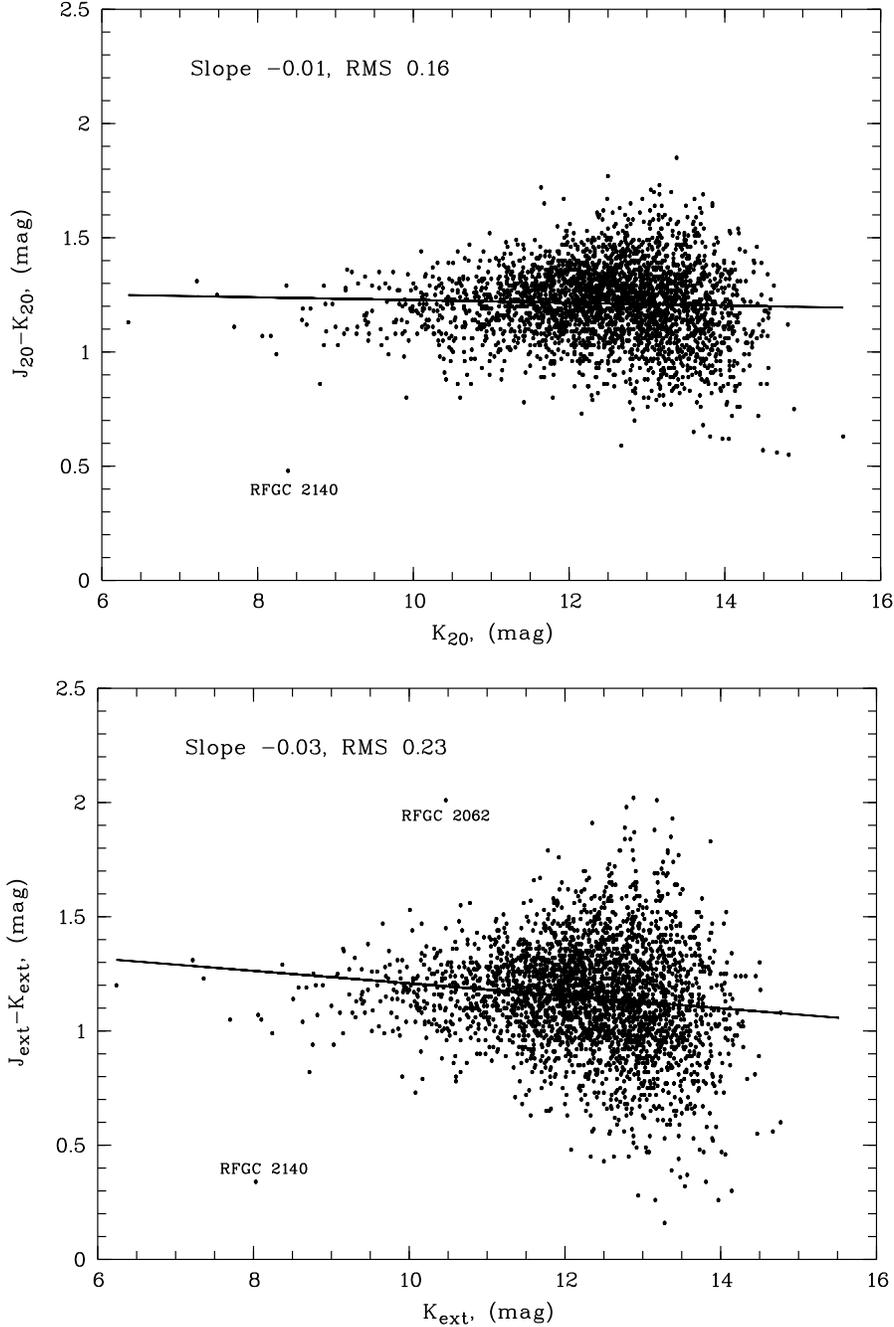

\centering
\includegraphics[width=9cm,angle=-90]{2893.F5a}
\includegraphics[width=9cm,angle=-90]{2893.F5b}
\vspace{5mm}
\caption{ Top: the $(J_{20} - K_{20})$ color vs.  apparent $K_{20}$
magnitude. Bottom: the  $(J_{ext} - K_{ext})$ color vs. the total apparent
magnitude $K_{ext}$.}
\end{figure*}

  Unfortunately, we did not find in literature a representative enough sample
of data on deep photometry for RFGC galaxies in the $J, H, K$ bands. Such a
kind of data are available in the $I$ band only; that is why our comparison
of `fast' 2MASS photometry with the photometry based on `deep' exposures
will be indirect. Mathewson \& Ford (1996) published an enormous list
of 2447 southern spiral galaxies observed by them in the $I$ band with
the 1-m and 2.3-m telescopes of Siding Spring Observatory. Later
Haynes et al. (1999) presented their results of CCD photometry of 1727
spiral galaxies of Sbc and Sc types in the $I$ band. The observations
were carried out on three different telescopes with typical exposures
$\sim$600 sec, which enabled photometrying down to isophote $I\sim23.5^m/\sq\arcsec$.
A comparison of these two extended photometric sets shows that the integrated
I magnitudes of galaxies have been determined with an error of $\sim0\fm04$.
 In the lists (Mathewson \& Ford 1996, and Haynes et al. 1999) we found
479 RFGC galaxies detected in 2MASS.

\begin{figure*}
\centering
\includegraphics[width=9cm,angle=-90]{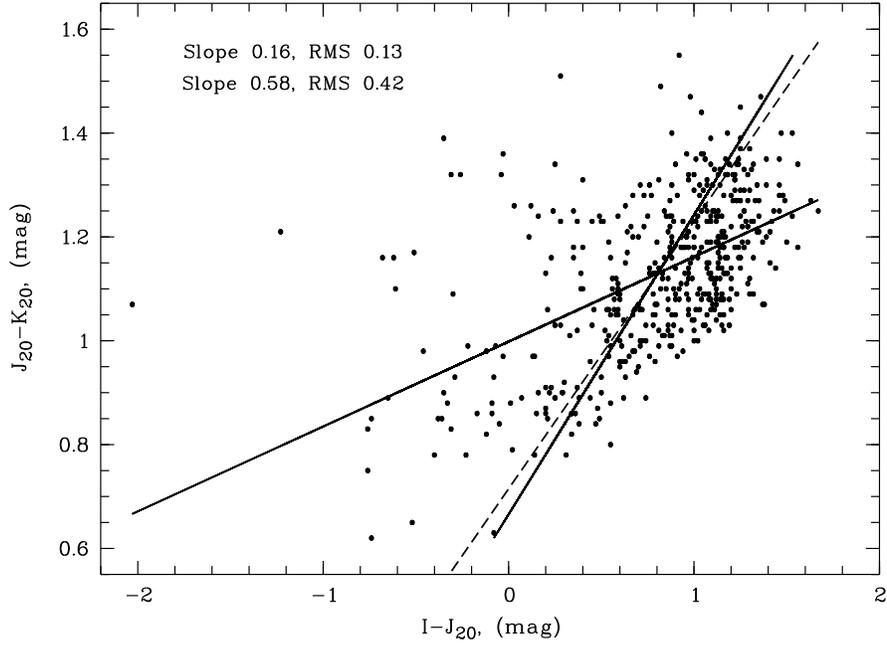}
\vspace{5mm}
\caption{  $(J_{20} - K_{20})$ color vs.  $(I - J_{20})$ color for
the edge-on galaxies. Two solid lines correspond to the regression lines.
The dashed line indicates the internal selective extinction to be the same
as in the Milky Way.}
\end{figure*}

\begin{figure*}
\centering
\includegraphics[width=9cm,angle=-90]{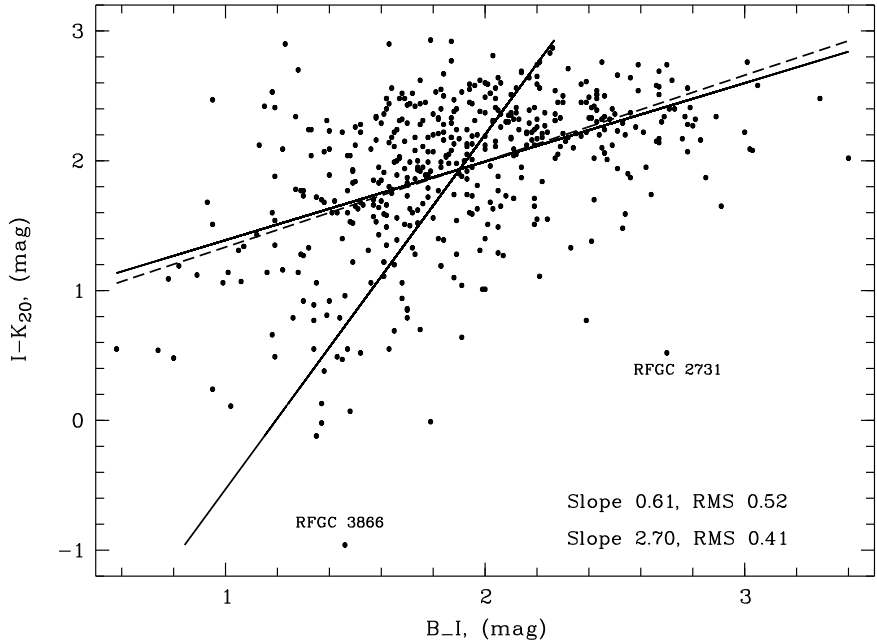}
\vspace{5mm}
\caption{  \{$I-K_{20}$ vs. $B-I$\} color - color diagram for the RFGC galaxies.
Two solid lines correspond to the regression lines. The dashed line
indicates the internal selective extinction to be the same as in the Milky
Way.}
\end{figure*}

The distribution of these galaxies on the `color-color' diagrams is given in
Figures 6 and 7. The first one corresponds to a `short' color scale,
$(J_{20} - K_{20})$ vs. $(I - J_{20})$. The galaxy distribution is characterized
by direct and inverse regressions having slopes of 0.16 and 0.58,
respectively, and the standard deviations $\sigma(J_{20}-K_{20}) = 0\fm13$ and
$\sigma(I-J_{20}) = 0\fm42$. The color difference between spiral galaxies is
mostly due to their difference in stellar population. However, in the
case of edge-on galaxies the different internal extinction comes to be
the main cause of the color difference. According to Giovanelli
et al. (1994) and Tully et al. (1998), the larger the linear dimension of a
spiral galaxy, the stronger the internal extinction is. Blue dwarf
galaxies of Sm type are almost transparent systems, while in giant
edge-on Sb,Sc galaxies their total optical luminosity dims several times.
Adopting the selective extinction in other galaxies to be the same as in the
Milky Way (Cardelli et al. 1989), we get the selectivity line with a
slope of 0.52, drawn in Fig.6 with a dashed stright line. As it is seen,
the selective extinction line fits well the observed behaviour of
edge-on galaxies in the \{$J_{20}-K_{20}$ vs. $I-J_{20}$\} diagram. The largest deviations
from the regression lines are characteristic of blue $(I-J_{20} <$ 0) galaxies
of low surface brightness, whose isophotal $J_{20}$ magnitudes are
apparently underestimated with respect to their total magnitudes.
In spite of the presence of the `blue wing', the \{$J_{20}-K_{20}$ vs. $I-J_{20}$\} diagram shows
that the mean external error of the $J_{20}$ magnitudes does not exceed
0$\fm$40, being suitable for mass measurements of distances of galaxies
via the infra-red TF relation.

  Using the $B$ magnitudes from RFGC for these 479 galaxies, we plotted in
Fig. 7 the `color-color' diagram with a maximum difference in wavelengths.
The distribution of edge-on galaxies in the \{$I-K_{20}$ vs. $B-I$\} diagram is
described by the direct and inverse regression lines having slopes of 0.61
and 2.70, and a RMS scatters of 0$\fm$52 and 0.41 mag, respectively.
The dashed line of `natural' extinction with a slope of 0.66 fits well the
observed distribution of RFGC galaxies, if one takes into account that the
typical error of B magnitudes is $\sim0\fm3$, and also the fact that
$K_{20}$ magnitudes for blue galaxies of low surface brightness are
systematically underestimated.

\section{ Tully-Fisher relation in different bands}

  As it has been noted by many authors, the slope of the TF relation increases,
but the scatter of galaxies decreases from blue to infra-red wavelengths.
This can be understood as an input of the young blue stellar population,
which only slightly affects the total luminosity of spiral galaxies in NIR
bands. For edge-on galaxies their internal selective extinction is an
additional source of differences of the TF diagrams plotted in different
bands.

\begin{figure*}
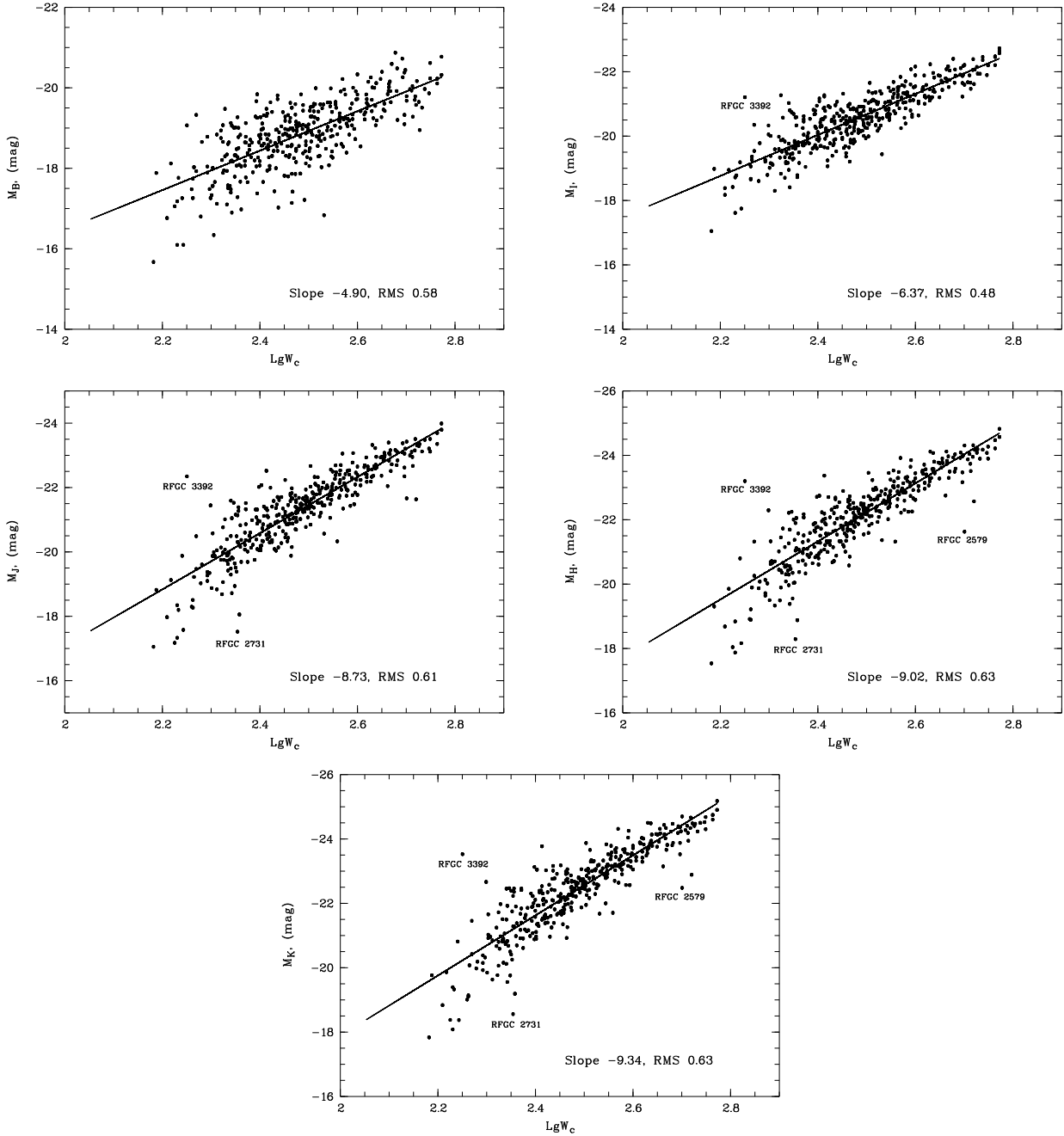

\centering
\includegraphics[width=5.9cm,angle=-90]{2893.F8a}
\includegraphics[width=5.9cm,angle=-90]{2893.F8b}
\includegraphics[width=5.9cm,angle=-90]{2893.F8c}
\includegraphics[width=5.9cm,angle=-90]{2893.F8d}
\includegraphics[width=5.9cm,angle=-90]{2893.F8e}
\vspace{5mm}
\caption{ Tully-Fisher relation for the RFGC galaxies in different
photometric bands.}
\end{figure*}

  Among of 479 RFGC galaxies with know $B,I,J,H$, and $K$ magnitudes there are
450 galaxies with the HI line width, $W_c$, measured at a 50\% level of the
maximum (Karachentsev et al. 2000). The absolute magnitude in different bands
for them was calculated as
      $$ M_\lambda = m_\lambda - A_\lambda - 25 -5 log(V_{LG}/H_0),$$
where the distance of a galaxy was determined from its radial velocity with
respect to the Local group centroid (Karachentsev \& Makarov, 1996),
adopting the Hubble constant $H_0 = 75~km~s^{-1}~Mpc^{-1}$.
The correction for the Galactic extinction, $A_\lambda$, was taken from
Schlegel et al.(1998). Plotting relations  $M$ vs. $log(W_c)$, we omitted
14 dwarf galaxies with $W_c <$ 150 km~s$^{-1}$ because of uncertainty of
their narrow HI line widths. The distribution of the
remaining 436 RFGC galaxies in the TF diagrams from the $B$ to
$K$ bands is shown in 5 panels of Fig. 8. In Table 1 we present the
slope of the direct TF regression in each band, as well as the RMS scatter
of galaxies with respect to it. For a reference, Table 1 gives also the
TF parameters when the total magnitudes ($J_{exp}, H_{exp}, K_{exp}$) or
the Kron magnitudes ($J_{fe}, H_{fe}, K_{fe}$) are used instead of the
isophotal ones. (Kron aperture photometry (Kron 1980) employs a method
in which the aperture is adapted to the first image moment radius).
Examination of these data allows us to note some properties of the
TF diagrams for RFGC galaxies.

\begin{table}[h]
\caption{Statistical properties of the TF-relations for 436 RFGC galaxies}
\begin{tabular}{ccc}  \hline
	   Photometric  & direct regression  &  RMS scatter   \\
	      band      &     slope          &    (mag)       \\ \hline
	       B        &   $-$4.90$\pm$0.21       &    0.58  \\
	       I        &   $-$6.37$\pm$0.18       &    0.48  \\
	 J$_{20}$       &   $-$8.73$\pm$0.24       &    0.61  \\
	 J$_{ext}$      &   $-$8.20$\pm$0.24       &    0.68  \\
	 J$_{fe}$       &   $-$7.74$\pm$0.23       &    0.66  \\
	 H$_{20}$       &   $-$9.02$\pm$0.25       &    0.63  \\
	 H$_{ext}$      &   $-$8.71$\pm$0.25       &    0.73  \\
	 H$_{fe}$       &   $-$7.95$\pm$0.23       &    0.66  \\
	 K$_{20}$       &   $-$9.34$\pm$0.25       &    0.63  \\
	 K$_{ext}$      &   $-$9.02$\pm$0.25       &    0.71  \\
	 K$_{fe}$       &   $-$8.38$\pm$0.24       &    0.69  \\

\hline
\end{tabular}
\end{table}

\begin{figure*}
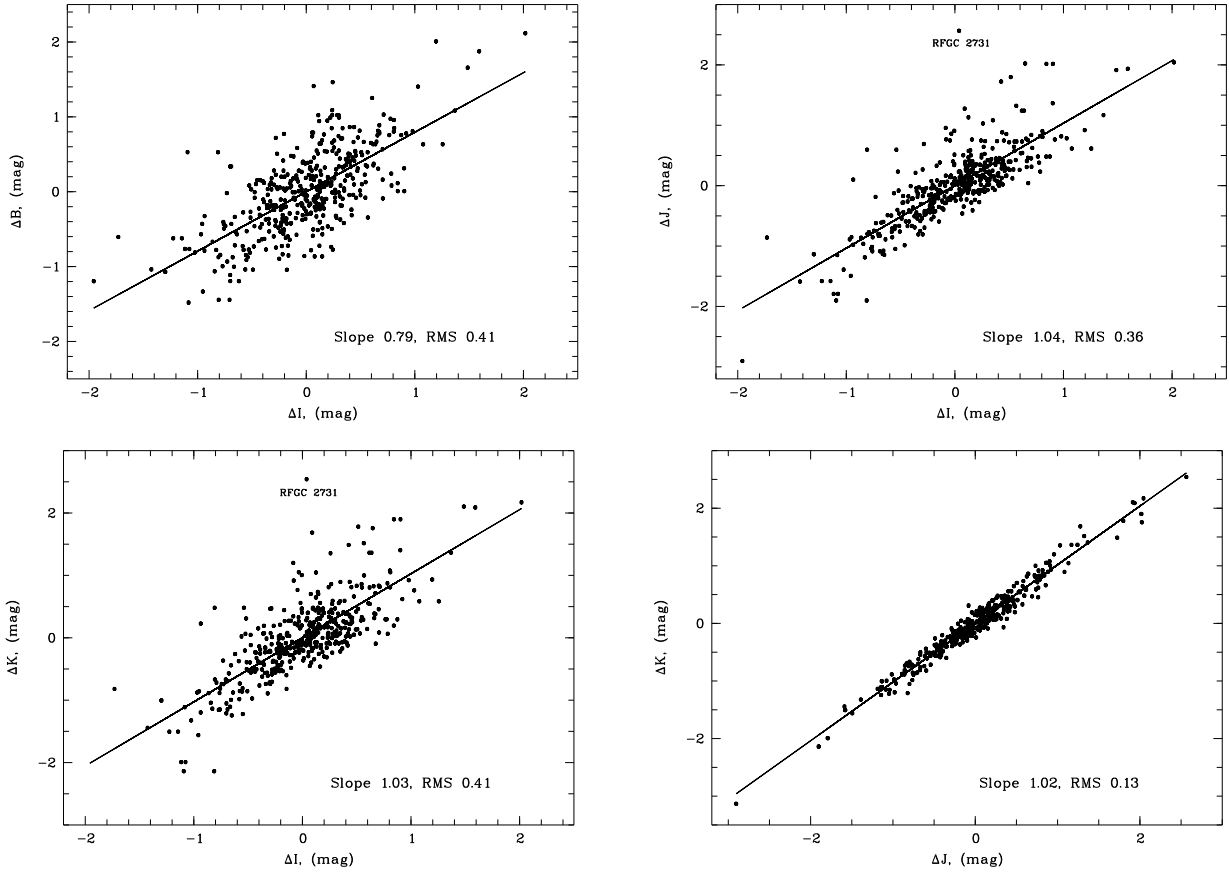

\centering
\includegraphics[width=5.9cm,angle=-90]{2893.F9a}
\includegraphics[width=5.9cm,angle=-90]{2893.F9b}
\includegraphics[width=5.9cm,angle=-90]{2893.F9c}
\includegraphics[width=5.9cm,angle=-90]{2893.F9d}
\vspace{5mm}
\caption{ Residual distributions with respect to the TF regression line in
different photometric bands.}
\end{figure*}

  The absolute value of the slope of the linear regression increases steadily
with increasing a wavelength of the photometric band. The slope difference
from 4.9 in the $B$ band to 9.3 in the $K$ band turns out to be much higher
than for spiral galaxies of arbitrary orientation. For example, for
spiral galaxies in the Ursa Majoris cluster the slope increases only
from 6.8 in the $B$ band to 8.0 in the $K$ one (Verheijen, 2001). This
difference can be readily explained if dwarf galaxies $(log W_c \simeq 2.2)$
are almost transparent systems, but giant spirals $(log W_c \simeq 2.75)$ have
the mean internal extinction  $<\Delta B> = 1\fm8$. About the same
value, $\Delta B = 1\fm7$, has been derived by Tully et al. (1998) for
the most luminous edge-on galaxies in Ursa Majoris.

  Fig. 8 and Table 1 do not display any significant decrease in the
scatter of galaxies in TF diagram from optical bands to NIR. The tendency
of $\sigma(M)$ to decrease with $\lambda$ is seen for the most luminous
galaxies only. This circumstance is not a trivial one because it is the
brightest spiral galaxies that are subjected to strong internal extinction,
dimming, on the average, their blue luminosity by a factor of 5. Another
explanation of the $\sigma(M)$ vs. $\lambda$ dependence may be related
with uncertainties in HI line widths. For instance, a scatter of
$\sigma(log W_c)$ = 0.05 generates a magnitude error of 0.25 mag in the
$B$ band and 0.47 mag in the $K_{20}$ band.

  As it is obvious, the errors of photometry done in the $J,H,K$ bands are
independent of the photometry errors in the $B$ and $I$ bands.
Nevertheless, the residuals of galaxies in the TF diagrams exibit tight
mutual correlations. Fig. 9 shows the distributions of 436 RFGC galaxies
according to their residuals with respect to the regression lines in the $B,I,
J$, and $K$ bands. The greatest scatter (RMS = 0$\fm41$) is seen on the
\{$\Delta B$ vs. $\Delta I$\} diagram, which is due to errors of determination of
$B$ magnitudes ($\sim0\fm3$) via the angular diameter and surface brightness of
the galaxies. In the diagrams \{$\Delta J$ vs. $\Delta I$\} and \{$\Delta K$ vs. $\Delta I$ \}
the galaxy distributions look rather asymmetric because the $J_{20}$
\begin{figure*}[hbt]
\centering
\includegraphics[width=9cm,angle=-90]{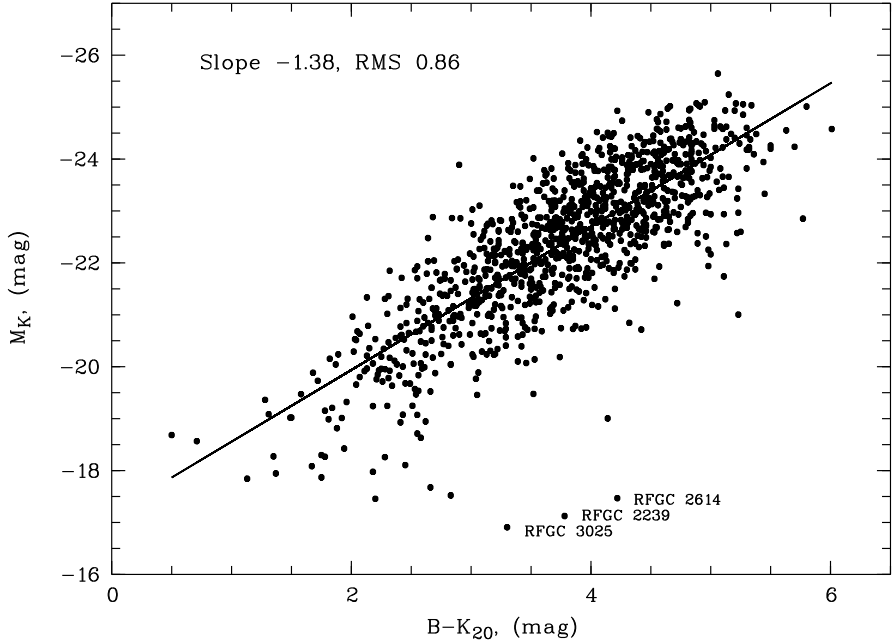}
\vspace{5mm}
\caption{ Absolute $K_{20}$- magnitude vs. $(B-K_{20})$ color for the edge-on
galaxies. The regression line has a slope of -1.38. Three the most deviated
galaxies are indicated by their RFGC numbers.}
\end{figure*}
and $K_{20}$ magnitudes underestimate the total luminosity of galaxies having
low surface brightness. In the diagram \{$\Delta K$ vs. $\Delta J$\} aperture
corrections are mutually dependent, yielding a scatter of 0$\fm$13 only.

  The observed correlation of residuals of the galaxies relative to the
TF regressions shows that the galaxy scatter is mainly caused by some
physical/statistical reasons, but not by photometric errors. Such reasons may
be: a) large non-Hubble motions of galaxies, b) differences in structure of
galaxies, particularly, in their visible mass-to-dark mass ratio,
c) errors in measurements of HI line widths.

\section {Color-luminosity relation for edge-on galaxies}

As it follows from the above- presented data, the spiral galaxies seen
edge-on have a vast diversity of colors. For instance, their $B-J$ and
$B-K$ colors span a range of 5 magnitudes! The bluest edge-on galaxies
belong to almost transparent dwarf systems, where the young blue stellar
population predominates. The reddest edge-on galaxies have giant disks
whose dusty component dims their blue flux $\sim$5 times. Because the
galaxy color (after the K-correction) does not depend on its distance,
the `absolute magnitude - color' relation may be used to determine
distances of edge-on galaxies in the same manner as the classical TF
relation is used. Such a possibility has already been discussed by
Visvanathan (1981), Tully et al. (1982), and Kraan-Korteweg et al. (1988).
In Fig. 10 we present the distribution of 1100 RFGC galaxies by
their absolute magnitude and color, \{$M_K$ vs. $B-K$\}. The diagram shows
a close correlation between observables with a scatter of 0$\fm86$,
which is a little worse than for the usual TF relation. Usage of the
color --- luminosity relation offers a possibility of mass measurements
of distances to edge-on galaxies, based exclusively on photometric
surveys of the sky, like 2MASS, SDSS, etc. We believe that the accuracy
of the `color' distances can be improved by taking account of surface
brightness and other global parameters of the galaxies.

\section {Conclusions}

  We considered the 2MASS photometric data on galaxies from the RFGC catalog
with blue angular diameters $\geq 0\farcm6$ and apparent axial ratios
$a/b \geq 7$. About 71\% of the RFGC objects are seen in the $J,H$, and $K_s$
bands of 2MASS. Due to the short exposures, 2MASS does not detect the
faint periphery of the disks. As a result, in the infra-red bands the edge-on
galaxies seem to be about twice as short and also about two times as
round as in the optical $B$ band. The measured isophotal $J_{20}, H_{20},$
and $K_{20}$ magnitudes need an average correction of 0$\fm24$ to get the
total magnitudes. The mean internal error of the 2MASS photometry
is $\sim0\fm16$, allowing the $J,H$, and $K$ magnitudes to be used for
plotting the Tully-Fisher relation. The TF diagram in the $B,I,J,H$, and
$K$ bands has a typical scatter of 0$\fm5-0\fm6$, which can be appreciably
improved, by making allowance for surface brightness and other
parameters of edge-on galaxies. The slope of derived TF relations
increases steadily from 4.9 in the $B$ band till 9.3 in the $K$ one. The
effect is mainly due to the internal extinction in the galaxies seen edge-on.
The observed slope difference can be easily understood if blue dwarf
galaxies are practically transparent and giant galaxies with typical rotation
velocities of $W_c\sim550$ km/s have internal extinction of $\sim1\fm8$
in blue light. Being averaged over giant and dwarf galaxies, the mean
internal extinction, $\Delta B \sim 1\fm0$, well agrees with other estimates
made by Haynes \& Giovanelli (1984), Karachentsev (1991), and Verheijen
\& Sancisi (2001). On the color-color diagrams, $J-K$ vs. $I-J$ and
$I-K$ vs. $B-I$, the RFGC galaxies follow approximately
the line of selective extinction
typical of the Milky Way. The difference in stellar populations and,
especially, in internal extinction between dwarf and giant galaxies
gives rise to initiate correlation between color and luminosity for edge-on galaxies.
The moderate scatter of the RFGC galaxies in the `color-luminosity'
diagram, 0$\fm$86, provides us with a new approach to mass measurements
of distances to galaxies on the basis of modern photometric sky surveys
like 2MASS and SDSS.

 \acknowledgements{ We thank the referee, G. Paturel, for his very useful
comments. This paper makes use of data of the the 2MASS, which is a
joint project of the University of Massachusetts and the Infrared
Processing and Analysis Center, funded by the NASA and NSF.
This research was partially supported by DFG-RFBR grant 436 RUS 113/701/0-1.}

{}


\begin{thebibliography}{}

\bibitem {} Cardelli J.A., Clayton G.C., Mathis J.S., 1989, ApJ, 345, 245
\bibitem {}de Vaucouleurs G., de Vaucouleurs A., Corwin H.G. et al., 1991, Third
	     Reference Catalogue of Bright Galaxies, Berlin, Springer (RC3)
\bibitem {}Giovanelli R., Haynes M.P., Salzer J., et al., 1994, AJ, 107, 2036
\bibitem {}Giovanelli R., Avera E., Karachentsev I.D., 1997, AJ, 114, 122
\bibitem {}Haynes M.P., Giovanelli R., 1984, AJ, 89, 758
\bibitem {}Haynes M.P., Giovanelli R., Chamaraux P., et al.,1999, AJ, 117, 2039
\bibitem {}Jarrett T.H., 2000, PASP, 112, 1008
\bibitem {}Jarrett T.H., Chester T., Cutri R., et al., 2002, AJ, submitted
\bibitem {}Jarrett T.H., Chester T., Cutri R., et al., 2000, AJ, 119, 2498
\bibitem {}Karachentsev I., 1989, AJ, 97, 1566
\bibitem {}Karachentsev I., 1991, Letters to Astron. Zh., 17, 671
\bibitem {}Karachentsev I.D., Makarov D.I., 1996, AJ 111, 535
\bibitem {}Karachentsev I.D., Karachentseva V.E., Parnovsky S.L., 1993, AN, 314,
 97 (FGC)
\bibitem {}Karachentsev I.D., Karachentseva V.E., Kudrya Yu.N., Makarov D.I.,
    2000, Bull. SAO, 50, 5
\bibitem {}Karachentsev I.D., Karachentseva V.E., Kudrya Yu.N., et al., 1999,
	     Bull. SAO, 47, 5 (RFGC)
\bibitem {}Kraan-Korteweg R.C., Cameron L.M., Tammann G.A., 1988, ApJ 331, 620
\bibitem {}Kron R.G., 1980, ApJS 43, 305
\bibitem {}Kudrya Yu.N., Karachentseva V.E., Karachentsev I.D., et al., 1997,
	 Letters to Astron. Zh., 23, 728
\bibitem {}Mathewson D.S., Ford V.L., 1996, AJ, 107, 97
\bibitem {}Nikolaev S., Weinberg M.D., Skrutskie M.F. et al., 2000, AJ, 120, 3340
\bibitem {}Schlegel D.J., Finkbeiner D.P., Davis M., 1998, ApJ, 500, 525
\bibitem {}Tully R.B., Fisher J.R., 1977, A\&A, 54, 661
\bibitem {}Tully R.B., Mould J.R., Aaronson M., 1982, ApJ, 257, 527
\bibitem {}Tully R.B., Pierce J., Huang J.S., et al., 1998, AJ, 115, 2264
\bibitem {}Verheijen M.A., 2001, ApJ, 563, 694
\bibitem {}Verheijen M.A., Sancizi R. 2001, A\&A, 370, 765
\bibitem {}Visvanathan N., 1981, A\&A 100, L20
\end{thebibliography}
\end{document}